%Paper: hep-th/9303094
%From: Ken-ichiro Aoki <aoki@madonna.physics.ucla.edu>
%Date: Tue, 16 Mar 93 18:51:46 -0800

\input phyzzx
\def\oxi{\bar \xi}
\def\oth{\bar \theta}
\def\d{{\cal D}}
\def\f{{\cal F}}

\def\n{{\cal N}}

\def\eg{{e.g.}}

\def\gm{{\hat g_{mn}}}
\def\half{{1 \over 2}}
\def\sc{\quad;\quad}
\def\oddroots{\Delta_s^{\rm odd}}
\def\evenroots{\Delta_s^{\rm even}}
\def\eg{{\it e.g.}}
\def\nl{\hfil\break}
\def\anp{{ \sl Ann. Phys. }}
\def\cmp{{ \sl Comm. Math. Phys. }}
\def\ijp{{ \sl Int. J. of Mod. Phys. }}

\def\lmp{{ \sl Lett. Math. Phys. }}

\def\npb{{ \sl Nucl. Phys. }}

\def\prd{{ \sl Phys. Rev. }}
\def\prl{{ \sl Phys. Rev. Lett. }}
\def\plb{{ \sl Phys. Lett. }}
\def\rmp{{ \sl Rev. Mod. Phys. }}

\def\c#1{{\cal{#1}}}
\def\inbar{\,\vrule height1.5ex width.4pt depth0pt}
\def\IC{\relax\hbox{$\inbar\kern-.3em{\rm C}$}}
\def\IR{\relax{\rm I\kern-.18em R}}
\font\cmss=cmss10 \font\cmsss=cmss10 at 7pt
\def\IZ{\relax\ifmmode\mathchoice
{\hbox{\cmss Z\kern-.4em Z}}{\hbox{\cmss Z\kern-.4em Z}}
{\lower.9pt\hbox{\cmsss Z\kern-.4em Z}}
{\lower1.2pt\hbox{\cmsss Z\kern-.4em Z}}\else{\cmss Z\kern-.4em Z}\fi}
\overfullrule=0pt
\Pubnum={UCLA/92/TEP/36}
\date={}
\titlepage
\title{Geometrical origin of integrability
for Liouville and Toda theory\foot{
Invited  talk presented at the 877th meeting of the American
Mathematical Society, USC, November 1992 and at the YITP
workshop ``Directions on Quantum Gravity", Kyoto, November
1992.
Research supported in part by the National Science Foundation
 grant NSF--PHY--89--15286.}}
\author{\rm Kenichiro Aoki and
Eric D'Hoker\foot{Electronic mail addresses: {\tt aoki@physics.ucla.edu,
dhoker@uclahep.bitnet.}}}
\def\ucla{Department of Physics\break
      University of California Los Angeles\break
        Los Angeles, California 90024--1547}
\address{\ucla}
\abstract{
We generalize the Lax pair and B\"acklund transformations for Liouville
and Toda field theories as well as their supersymmetric generalizations,
to the case of arbitrary Riemann surfaces.
We make use of the fact that Toda field theory arises naturally and
geometrically in a restriction of so called $W$--geometry to ordinary
Riemannian geometry.
This derivation sheds light on the geometrical structure underlying
complete integrability of these systems.
}
\endpage
\singlespace
\noindent
{\bf 1.  Introduction.}

It has long been known that certain $1+1$ dimensional classical field
theories are completely integrable.  This property has been characterized
in a wide variety of ways, including the existence of an infinite number
of conserved charges, the existence of a B\"acklund transformation, the
existence of a Lax pair, the solvability by the inverse scattering
method and so on.
\footnote\dagger{For some of the original work
and standard reviews, see [1].}
Perhaps the most fundamental
characterization of all is that the field equations for the system arise as
the flatness condition on a certain connection or gauge field.
If this is so, then a Lax pair may always be deduced, and from it
a B\"acklund transformation, the inverse scattering solution and infinite
numbers of conserved charges.  Given an equation though, it is in
general far from clear how to conclude whether a given system is a
flatness condition on some connection.  Many attempts at finding such an
algorithm have been made, but at present it is unclear that any useful
procedure indeed exists.

Short of a decisive test on complete integrability, one may proceed from
the opposite direction, and trace back the existence of a Lax pair to
the geometrical context in which the completely integrable system
arises.  It has long been suggested that all these integrable systems
are special cases of the self--duality equations on four dimensional
gauge fields, equations that are known to be completely integrable [2].
Beautiful as this connection may be, it is also perhaps too general, and
gives little clue as to why specifically any system arises as a
reduction of the self--duality equations.

In some recent work, it was shown that Liouville theory [3] on a general
background geometry, arises as a constant curvature condition, which in
turn is related to a flatness condition on an $SL(2,{\IR})$ connection
[4] as discovered in the group manifold approach [5], as well as in
topological field theory considerations [6,7].  The corresponding Lax
pair is given by the parallel transport equation under this
$SL(2,{\IR})$ connection [4].  This derivation is easily generalized to
the case of $N=1$ super--Liouville theory, where the relevant gauge group
is $OSp(1,1)$ [4].

A natural extension to include the case of Toda field theory coupled to
an arbitrary background geometry, requires the extension of
2--dimensional Riemannian geometry to $W$--geometry [8],
in which in addition to the
spin 2 metric, additional higher spin fields are coupled.  Toda field
theory, here arises as a natural reduction of general $W$--geometry to
ordinary 2--d Riemannian
geometry, and the flat connection naturally arises as the
Maurer--Cartan form on higher rank groups, generalizing $SL(2,{\IR})$, or its
supersymmetric generalizations [9].

We shall review and slightly extend the above results in these lectures.

\vfill\eject

\noindent
{\bf 2.  Liouville Theory and 2--d Riemannian Geometry}

The classical Liouville equation, on an arbitrary background geometry,
determines the Weyl factor that scales an arbitrary 2--d metric to a
constant curvature metric; and follows from the Weyl transformations
properties of the curvature.  The Liouville equation
$$
\Delta_{\hat g} \phi + R_{\hat g} + \mu^2 e^{2\phi} = 0 \eqno(2.1)
$$
is equivalent to
$$
R_g + \mu^2 = 0 \qquad\qquad g_{mn} = e^{2\phi} \gm \eqno(2.2)
$$
where $\mu^2$ is a real constant.

Two dimensional geometry is parameterized by
\footnote\dagger{Einstein indices are denoted by $m,n,\cdots$ and frame
indices by $a,b\cdots$.  Frame indices split up into complex
conjugates under $U(1)$ frame rotations $a = (z,\bar z)$ where
$\delta_{z\bar z} = \delta_{\bar z z} = 1$, $\epsilon_z{}^z = -
\epsilon_{\bar z}{}^{\bar z} = i$.  The metric is given by $g_{mn}$ =
$e_m{}^a e_n{}^b \delta_{ab}$ and the Gaussian curvature by $R_g= \half
\epsilon^{mn} R_{mn}$.}
a frame (or zweibein) $e^a = d\xi^me_m {}^a $ and a $U(1)$ spin
connection $\omega = d\xi^m\omega_m $.  Torsion and curvature are
defined by
$$
\eqalign{
T^a & = d e^a + e^b \wedge \omega \epsilon_b{}^a = \half d\xi^n
\wedge d\xi^m T_{mn}{}^a\cr
R & = d\omega = \half d\xi^n \wedge d\xi^m R_{mn} \cr}
\eqno(2.3)
$$
Covariant derivatives acting on tensors of weight $n$ are defined by
$$
D_a{}^{(n)} \equiv e_a{}^m (\partial_m + i n \omega_m)
$$
so that the Laplacian on scalars takes the form
$$
\Delta_g = -2 D_{\bar z} D_z{}^{(0)}
$$
Weyl transformations are defined by
$$
e_m{}^a = e^\phi \hat e_m{}^a, \qquad \omega_m = \hat\omega_m +
\epsilon_m{}^p \partial_p\phi\eqno(2.4)
$$
under which torsion and curvature transform as
$$
T^a = e^\phi \hat T^a \qquad\qquad R_g = R_{\hat g}e^{-2\phi} + \Delta_g
\phi \eqno(2.5)
$$
Zero torsion and constant curvature are equivalent to Liouville's
equation (2.1), as can be seen from (2.2).  We now show that these
condition arise as a flatness condition on an $SL(2,{\IR})$ valued
connection.

Consider the algebra $G$, with generators $J_3, J_z$ and $J_{\bar z}$,
satisfying
$$
[J_3, J_z] = J_z \quad ; \quad [J_3, J_{\bar z}] = - J_{\bar z};\qquad
[J_z, J_{\bar z}] = - 2 \lambda_z \lambda_{\bar z} J_3 \eqno(2.6)
$$
The fundamental representation can be written in terms of
$2 \times 2$ Pauli matrices:
$$
J_3 = \half \sigma^3
;\qquad J_z = {\lambda_z \over 2} (\sigma^1 +
i\sigma^2) ;\qquad
J_{\bar z} = {\lambda_{\bar z} \over 2} (-\sigma^1 +
i\sigma^2) \eqno(2.7)
$$
We define a $G$--valued gauge field, constructed from the frame $e^a$ and
the $U(1)$ spin connection $\omega$ by assembling these fields as follows
$$
\c A = -i\omega J_3 + e^z J_z + e^{\bar z} J_{\bar z} \eqno(2.8)
$$
The  curvature form is easily evaluated
$$
\c F  = d\c A + \c A \wedge \c A
 =  ( -iR-2\lambda_z\lambda_{\bar z} e^z \wedge e^{\bar z} )
 J_3 + T^z J_z + T^{\bar z} J_{\bar z}\eqno(2.9)
$$
Setting $\c F = 0$ yields $T^a = 0$ and constant curvature as in (2.2) with
$\mu^2 = -2 \lambda_z \lambda_{\bar z}$.  Notice that this algebra in
$SU(2)$ for $\mu^2 < 0, SL(2,{\IR})$ for $\mu^2 > 0$ and the
2--dimensional Euclidean group $E(2)$ for $\mu^2 = 0$.

Geometrically, the significance of the flatness condition $\c F=0$ is as
follows.  When $\c F=0$, $\c A$ is the Maurer-Cartan form on $G$, parameterized
by the underlying Riemann surface.  This form always exists globally on
the surface.  Upon projection onto the symmetric space $G/U(1)$, the
$U(1)$ generator becomes the spin connection on the coset space, and the
generators complementary to $U(1)$ become the frame on the coset space.
In fact, more generally, higher dimensional conditions for symmetric
spaces can be expressed as flatness condition as well [10].

By construction, a flatness condition $\c F=0$ is the integrability
condition on a system of first order differential equations, namely the
equations for parallel transport.  This provides us
right away with the correct expressions for the Lax pair.  Introducing
\eg\ a $G$--doublet field $\psi$, we consider the system of first order
linear differential equations in $\psi$
$$
\left ( \partial_m + \c A_m\right ) \psi = 0 \qquad\qquad \psi = \pmatrix{
\psi_1\cr
\psi_2\cr} \eqno(2.10)
$$
where $\c A_m$ is given in (2.8).  This system is integrable when $\c F_{mn} =
0$, which, as shown above, is precisely equivalent to Liouville's
equation.  This can be seen even more clearly by first performing the
Weyl rescaling (2.4) on (2.8), yielding the following Lax pair for
Liouville theory on a general 2--d background geometry, given by frame
$\hat e_m{}^a$ and $U(1)$ spin connection $\hat \omega_m$:
$$
\left( \partial_m - i \hat\omega_m J_3 - i\epsilon_m{}^p
\partial_p\phi J_3 + e^\phi \hat e_m{}^z J_z + e^\phi\hat
e_m{}^{\bar z} J_{\bar z}\right ) \psi=0 \eqno(2.11)
$$
By identification with the Lax pair for Toda field theory
on flat space-time, we see that
$\lambda = \lambda_z/\lambda_{\bar z}$
plays the role of a spectral parameter,
which is always real for real $\mu^2$.
This parameter emerges in a natural group theoretical way within this
context.

 From the Lax pair, we recover the B\"acklund transformation in a
well--known way by introducing the field $\sigma: \psi_2 = e^\sigma
\psi_1$ and we get
$$
\partial_m\sigma + i\hat\omega_m + i\epsilon_m{}^p \partial_p\phi =
\lambda_{\bar z} \hat e_m{}^{\bar z} e^{(\phi-\sigma)} +
\lambda_z\hat e_m{}^ze^{(\phi+\sigma)} \eqno(2.12)
$$
The integrability condition on this system viewed as an equation for
$\sigma$ is precisely the Liouville equation (2.1), and if viewed as an
equation for $\phi$, the integrability condition
is a linear equation in $\sigma$:
$$
\Delta_{\hat g}\sigma = \hat \epsilon^{mn} \partial_m (\hat\epsilon_n{}^p
\hat\omega_p)\eqno(2.13)
$$
The Lax pair and B\"acklund transformations reduce to the ones for flat
background geometry, and again allow for a complete explicit solution of
(2.1) [3].

The flatness condition, and the equation for parallel transport may be
derived from an action principle.
$$
S = \int \tr \n \c F\eqno(2.14)
$$
where $\n$ is an auxiliary field in the (co) adjoint representation of
the algebra.  Upon eliminating $\n^z$ and $\n^{\bar z}$, torsion is set
to zero, and the remaining action coincides with the one proposed for
2--d gravity by Jackiw and Teitelboim [6].

\bigskip

\noindent
{\bf 3.  Two--dimensional Dilaton Gravity.}

The $SL(2,{\IR})$ connection of (2.8) may be generalized by addition of
a field ``$a$'' multiplying a generator $I$ that commutes with all $J$'s
[11]:
$$
\c A = - i\omega J_3 + e^z J_z + e^{\bar z} J_{\bar z} + a I\eqno(3.1)
$$
and we postulate the following structure relations for $G$, satisfying
the Jacobi identity:
$$
[J_3, J_z] = J_z;\qquad
 [J_3, J_{\bar z}] = - J_{\bar z} ;\qquad [J_z, J_{\bar
z}] = \mu^2 J_3 + \lambda I ;\qquad [J_a, I] = 0  \eqno(3.2)
$$
For $\mu^2 \not= 0$, one may redefine the generator $J_3$ by $\mu^2
\tilde J_3 = \mu^2 J_3 + \lambda I$ so that the algebra is easily recognized
as $GL(2,{\IR})$ or $U(2)$ in the compact case.  However for the two
dimensional Poincar\'e algebra $(\mu^2 = 0)$, such a redefinition is
not possible, and we have a non--trivial central extension of this
Poincar\'e group.  The curvature of $\c A$ is:
$$
\c F = \left ( -iR + \mu^2 e^z \wedge e^{\bar z} \right ) J_3 + T^z J_z +
T^{\bar z} J_{\bar z} + \left (da + \lambda e^z \wedge e^{\bar z}
\right ) I \eqno(3.3)
$$
The field equations $\c F=0$ have the following interpretation:  for $\mu^2
= 0$, the interesting case, the geometry is flat $(R = 0)$
Riemannian $(T = 0)$ and the one form field ``$a$'' obeys an interesting
equation as well.  Under a Weyl transformation as in (2.4), these
equations reduce to those of $2$--dimensional dilaton gravity [12].
\bigskip

\noindent
{\bf 4.  Toda Field theory and Two--Dimensional $W$--geometry}

The study of $W$--geometry has begun only recently, and there are a
number of different formulations which are presumably equivalent.
$W$--geometry, also called $W$--gravity in physics literature,
 may be defined in terms of an action principle, in analogy
with ordinary two--dimensional gravity [7]
$$
S= \int \tr \n \c F \qquad\qquad \c F = d\c A + \c A
\wedge \c A \eqno(4.1)
$$
but now the fields $\c A, \c F$ and $\n$ assume values in a more general Lie
algebra $G$.

We define a Lie algebra $G$ (not necessarily finite dimensional) by the
following Chevalley relations
$$
[h_i, h_j] = 0, \qquad [h_i, x_{\pm\alpha_j}] = \pm k_{ij} x_{\pm
\alpha_{j}},\qquad [x_{\alpha_{i}}, x_{-\alpha_j} ] = \delta_{ij} h_i\eqno
(4.2)
$$
Here the generators of the Cartan subalgebra $H$ are denoted by $h_i$
and the positive (negative) simple roots by $x_{\alpha_{i}}$
$(x_{-\alpha_{i}})$ for $i=1,\cdots, r(\equiv$ rank$\,G)$.  To find all the
roots, one successively commutes the simple roots, using the constraints
of the Jacobi identity and the Serre relations
$$
Ad_{{x}_{\alpha_{i}}}^{(1-k_{ij})} x_{\alpha_{j}} = 0 \eqno(4.3)
$$
Equivalently, given the set of all roots $\Delta$, one has
$$
[x_\beta, x_\gamma] = N_{\beta\gamma} x_{\beta+\gamma} {\rm\  if\ }
x_{\beta+\gamma} \in \Delta \eqno(4.4)
$$
and $N_{\beta\gamma} = 0$ otherwise.  As a result, any root is a linear
combination of the $r$ simple roots, with integer coefficients which are
all positive (negative) for positive (negative) roots:  $\gamma =
\sum\limits_i \gamma^i \alpha_i$.  The height $\eta$ of a root is
defined by
$$
\eta(\gamma) = \sum_i \gamma^i \eqno(4.5)
$$
It vanishes on the Cartan subalgebra, and equals $1$ on any simple
(positive) root.

To define topological $W$--geometry,
we introduce a general $G$--valued connection
[13,9]
$$
\c A = \sum_i \omega^i h_i + \sum_{\gamma\in\Delta} e^\gamma x_\gamma
\eqno(4.6)
$$
Here,  $\omega^i$ are the components of the Abelian connection with gauge
group $H$, and $e^\gamma$ are the generalizations of the frame on the
Riemann surface.  As a generalization of ordinary Riemannian geometry,
$W$--geometry contains 2--d Riemannian geometry;
this is easily seen from the fact that any
semi--simple Lie algebra has an $SU(2)$ subalgebra.  The embedding of
2--d Riemannian geometry into $W$--geometry
is not unique in general, and has to be specified [13].
$$
e^{\alpha_i} = P_i e^z \quad ; \quad e^{-\alpha_i} = P_i e^{\bar
z}\quad ; \quad e^\gamma = 0{\rm\  if\ } |\eta(\gamma)|\geq 2\sc
P_i \left ( i \sum_j \omega^j k_{ji} - \rho_i \omega \right) = 0
\eqno(4.7)
$$
Under the maximal embedding $P_i = \rho_i = 1$ for all $i$, the spin of
the additional fields is simply related to the height of the
corresponding root by spin$(e^\gamma) = |\eta(\gamma)|+1$.
\footnote\dagger{The addition of $1$ results from the conversion of
Einstein indices into frame indices, analogous to isospin--spin
transmutation.}

The geometrical interpretation is that $\c A$ is a connection in the bundle
$G$ with structure group $H$ over the manifold $G/H$.  The latter is
always K\"ahler, so that the field contents of $W$--geometry may be
viewed as resulting from embedding a Riemann surface into a K\"ahler
manifold $G/H$, or equivalently from dimensional reduction of the
manifold $G/H$ to a Riemann surface.  Analogous embedding problems and
their relation to Toda systems were considered in [14].

The field strength of the connection $\c A$ may be recast in terms of the
frame field $e^\gamma$ and connection $\omega^i$,
$$
\c F  = \sum_i d\omega^i h_i + \half \sum_{\gamma\in\Delta} e^\gamma
\wedge e^{-\gamma} [x_\gamma, x_\gamma ]
+ \sum_{\gamma\in\Delta} \biggl [ de^\gamma +\sum_{i,j} \omega^i k_{ij}
\gamma^j \wedge e^\gamma + \half
\sum_{\scriptstyle\gamma',\gamma''\atop\scriptstyle\gamma'+\gamma''=\gamma}
e^{\gamma'} \wedge e^{\gamma''} N_{\gamma'\gamma''} \biggr]
\eqno(4.8)
$$
The first line on the right of (4.8) contains all the curvature terms in
the Cartan subalgebra, whereas in the second line we have the
contributions proportional to the roots.  The dynamics of $W$--gravity
governed by action (4.1) yields $\c F=0$, and this condition yields
constant curvature--type conditions from the generators in the Cartan
subalgebra, and torsion like conditions from the roots.
Note however that the torsion--type conditions are non--linear in the
frame fields.  $W$--geometry is analogous to supergeometry in which torsion
is not generally zero, but certain components are (covariantly)
constant.

Toda field theory corresponds to a very natural special case of the
general gauge field (4.6), in which the generators corresponding to
simple positive and negative roots only are retained, together with
those in the Cartan subalgebra, and all other are set to zero [9].  This
condition is a generalization of the maximal embedding of ordinary 2--d
Riemannian geometry but where now we allow for arbitrary complex scale factors
$\phi_i$:
$$
e^{\alpha_i} = \exp (\phi_i) e^z \qquad e^{-\alpha_i} = \exp
(\phi_i) e^{\bar z} \qquad
i \sum_j \omega^j k_{ji} = \omega \rho_i + e^a\epsilon_a{}^b D_b \phi_i
\eqno(4.9)
$$
for $i,j = 1, \cdots, r$ and $e^\gamma = 0$ when $|\eta(\gamma)| \geq
2$.  That this Ansatz is consistent can be shown from the expression for
the curvature in this case:
$$
\c F = \sum_i e^{\bar z} \wedge e^z \left [ - \left ( D_z\omega_{\bar
z}^i - D_{\bar z} \omega_z^i +
\exp (2\phi_i)-T_{z\bar z}{}^a\omega_a ^i\right )h_i
+ e^{\phi_i}T_{z\bar z}{}^z x_{\alpha_{i}} + e^{\phi_i} T_{z\bar
z}{}^{\bar z} x_{-\alpha_{i}}\right ]
\eqno(4.10)
$$
Thus for $r$ unknown fields $\phi_i$, we have $r$ equations,
together with the torsion constraints of the two dimensional background
geometry.  These $r$ fields satisfy the Toda field equations [15]
on a two dimensional geometry with frame $e^a$ and $U(1)$ spin
connection $\omega$:
$$
\Delta_g \phi_i + \sum_j \exp (2\phi_j) k_{ji} + R_g \rho_i = 0
\eqno (4.11)
$$
Recall that the Toda equations are Weyl or conformal invariant for any
underlying Lie algebra:
$$
g_{mn} \to g_{mn} e^{2\sigma};\qquad \phi_i \to \phi_i + \sigma
\eqno(4.12)
$$
However, when $G$ is a Kac--Moody algebra, the Cartan matrix has rank
$r-1$ and there is an eigenvector with zero eigenvalue, denoted by
$n^i$.  As a result, the particular combination $\phi = \sum\limits_i
n_i \phi^i$ satisfies a free field equation
$$
\Delta_g \phi + R_g \rho_i n^i = 0
\eqno(4.13)
$$
Elimination of this field results in breaking of the Weyl invariance for
the remaining equation.  In this way for example, one obtains the famous
sine--Gordon equation for $G = S\hat U(2)$.
To see this,  we define
$$
g_{mn} = e^{-\phi_1-\phi_2} \gm \quad ; \qquad \varphi = \half
(\phi_1-\phi_2)
\eqno(4.14)
$$
so that $\varphi$ satisfies a generalization of the sine-Gordon equation
to a general background geometry.
$$
\Delta_{\hat g}
\varphi + 4 \sinh2\varphi + R_{\hat g} = 0
\eqno(4.15)
$$
This equation is no longer Weyl invariant.

The Lax pair is identified as the equation for parallel transport under
the $G$--connection $\c A$ in some representation of $G$ [9]:
$$
\left( \partial_m + \c A_m\right ) \psi = 0
\eqno(4.16)
$$
In terms of frame index notation we get a set of very simple expressions
$$
\eqalign{
& \left ( D_z^{(0)} + \sum_i \omega_z^i h_i + \sum_i \exp (\phi_i)
x_{\alpha_{i}} \right ) \psi =   0\cr
& \left (D_{\bar z}^{(0)} + \sum_i \omega_{\bar z}^i h_i + \sum_i \exp (\phi_i)
x_{-\alpha_{i}} \right ) \psi = 0\cr}
\eqno(4.17)
$$
These equations now provide a Lax pair for Toda field theory on an
arbitrary Riemann surface.  Spectral parameters arise as in
Liouville theory.  For $G$ a Kac--Moody algebra, the field $\phi$ may be
eliminated from the Lax pair as well.

 From the Lax pair, we construct the B\"acklund transformation [9] by
passing from homogeneous coordinates $\psi$ of a linear representation
of $G$ to inhomogeneous coordinates of a non--linear realization of $G$.
We shall now examine how this is done for an arbitrary
(finite--dimensional) representation $\mu$ with highest weight vector
$\mu$, and highest weight $|0;\mu\rangle$ [16].  All other weights are built
by applying lowering operators:
$$
|j_1\cdots j_p;\mu \rangle = x_{-\alpha_{j_{p}}} \cdots x_{-\alpha_{j_{1}}}
|0;\mu\rangle
\eqno(4.18)
$$
where it is understood that $|j_1\cdots j_p;\mu\rangle = 0$
if the corresponding weight does not belong to the weight diagram of
$\mu$.  Application of Cartan generators and simple roots is
straightforward:
$$
\eqalign{
& h_j|j_1\cdots j_p ; \mu \rangle = \lambda_{j;\mu}^{(p+1)} |j_1\cdots j_p;\mu
\rangle\cr
& x_{-\alpha_{j}}|j_1\cdots j_p ; \mu \rangle = |j_1\cdots j_p j;\mu\rangle\cr
& x_{\alpha_{j}} | j_1\cdots j_p;\mu\rangle = \sum_{q=1}^p \delta_{j,j_{q}}
\lambda_{j;\mu}^{(q)} |j_1\cdots \hat j_q \cdots j_p;\mu \rangle \cr}
\eqno(4.19)
$$
Here the hat denotes omission and
$$
\lambda_{j;\mu}^{(q)} \equiv \mu_j - \sum_{m=1}^{q-1} k_{jj_{m}}
\eqno(4.20)
$$
B\"acklund conjugate variables are defined as
$$
\langle 0;\mu|\psi|0;\mu\rangle \exp \psi_ {\scriptstyle
{ {j_{1}}\cdots  \scriptstyle j_{p;\mu }}}
= \langle j_1\cdots j_p;\mu|\psi|0;\mu\rangle
\eqno(4.21)
$$
 From sandwiching the Lax equations between the states $\langle
j_1 \cdots j_p;\mu |$ and $|0;\mu\rangle$,
we get the B\"acklund transformations:
 $$
\eqalign{
&D_z ^{(0)}\left(\psi _{j_1 \dots j_p;\mu}-\sum_{q=1}^p \phi_{{j_q}}\right)
      +ip\omega_z + \sum _{i=1} ^r \left [
\exp \{ \phi _i + \psi _{j_1 \dots j_p i ;\mu}-\psi _{j_1 \dots j_p;\mu}\}
-\exp\{\phi_i+\psi_{i;\mu}\}
\right ] =0\cr
&D_{\bar z}^{(0)}
\left(\psi _{j_1 \dots j_p;\mu} +\sum_{q=1}^p\phi_{{j_q}}\right)
+ip\omega_{\overline z}
+\sum_{q=1}^p \lambda _{{j_q};\mu} ^{(q)} \exp \{
\phi _{j_q} + \psi _{j_1 \dots \hat j_q \dots j_p;\mu}-
\psi _{j_1 \dots j_p;\mu}\} =0 \cr }
 \eqno(4.22) $$
For $SU(n) = G$ and $\mu$ the fundamental representation, the rank of
the group is precisely the dimension of the fundamental representation
$-1$, so that the number of Toda fields $\phi_i$ and the number of
B\"acklund fields $\psi_i$ is the same.  In this case, we have a
B\"acklund transformation in the usual sense.

A few remarks are in order for the case of Kac--Moody algebras, when the
Cartan matrix has a zero eigenvalue.  In this case, equation (4.9) is
not invertible for the $U(1)$ connections $\omega^i$, and by the same
token its validity requires already a condition on the fields $\phi$.
Denote $n^i$ the zero eigenvector of $k$, then we must have
$$
0 = \omega \rho_i n^i + e^a\epsilon_a{}^b D_b \phi \eqno(4.23)
$$
where $\phi$ was defined in (4.13).  Equation (4.23) always implies
(4.13), but the reverse is only true in general on compact surfaces,
where the Laplacian is invertible up to a constant.  Furthermore,
condition (4.23) also imposes a restriction on the background geometry.
It is not in general true that $\omega$ is the curl of a single valued
scalar field $\phi$.  This requires that all its $1$--cycles are trivial
on the surface.  Of course, if $\phi$ is not assumed to be single
valued, then this restriction does not apply.

For the Kac--Moody case, there is another Lax pair realization, which we
shall  now discuss.
\footnote\dagger{A very similar construction was discussed in [17].}
We start with an ordinary Lie algebra $G$ of rank $r$, and Cartan matrix
$k_{ij}$ and we retain all the simple positive and negative roots, plus
the highest positive and negative roots, with all other roots zero:
$$
\eqalign{
& e^{\alpha_i} = e^z \exp \phi_i\cr
& e^\gamma = ie^{\bar z} \exp \phi\cr} \qquad\qquad
\eqalign{
& e^{-\alpha_i} = e^{\bar z} \exp \phi_i\cr
& e^{-\gamma} = ie^z \exp \phi\cr}
\eqno(4.24)
$$
where $\alpha_i$ are the simple positive roots $i = 1, \cdots, r$ and
$\gamma$ is the highest root: $\gamma= \sum\limits_i \gamma^i \alpha_i$.
With this Ansatz, the torsion equations in (4.8) are still linear in the
frame fields, and require
$$
i \sum_j \omega^j k_{ji} = \omega\rho_i + e^a \epsilon_a{}^b D_b
\phi_i \eqno(4.25a)
$$
$$
i \sum_{i,j} \omega^j k_{ji} \gamma^i = - \omega - e^a \epsilon_a{}^b D_b\phi
\eqno(4.25b)
$$
Here, since $k$ is the Cartan matrix of a finite dimensional algebra,
(4.25a) may always be solved for $\omega^j$.  Equation (4.25b) on the
other hand represents a constraint for the field $\phi$.

There are $r$ curvature equations for the fields $\phi_i$, which read:
$$
\Delta_g \phi_j - R_g \rho_j + \sum_i e^{2\phi_i} k_{ij} + \sum_i
e^{2\phi} \beta^i k_{ij} = 0
\eqno(4.26)
$$
where we have defined the vector $\beta$ from the highest root
commutator:
$$
\left[ x_\gamma , x_{-\gamma}\right] = \sum_i \beta^i h_i.\eqno(4.27)
$$
The first order differential equation (4.25b) implies a second order
equation, analogous to (4.26):
$$
\Delta_g \phi - R_g - \sum_{i,j} e^{2\phi_i} k_{ij} \gamma^j - \sum_{i,j}
e^{2\phi} \beta^i k_{ij} \gamma^j = 0
\eqno(4.28)
$$
Introducing the following $r+1$ dimensional quantities:
\def\rhoi{\rho_{\scriptscriptstyle I}}
\def\phii{\phi_{\scriptscriptstyle I}}
\def\phij{\phi_{\scriptscriptstyle J}}
$$
\eqalign{
\phii & = \left (\phii, \phi\right ) \qquad \rhoi=\left ( \rho_i,
1\right ) \qquad \beta^I = \left ( \beta^i, -1\right )\cr
k_{IJ} & = \left ({ k_{ij}\, \,\bigg |
-\sum\limits_{\ell} k_{i\ell} \gamma^\ell
\over
\sum\limits_{\ell} \beta^\ell k_{\ell j}
\,\bigg| -\sum\limits_{\ell,m} \beta^\ell
k_{\ell m}\gamma^m} \right )\cr}
\eqno(4.29)
$$
we may recast the Toda equation in the standard form
$$
\Delta_g \phii - R_g \rhoi + \sum_J e^{2\phij} k_{JI} = 0
\eqno(4.30)
$$
However, the Cartan matrix $k$ now has one eigenvector with zero
eigenvalue
$$
\sum_I \beta^I k_{IJ} = 0
$$
with all other eigenvalues positive.  In fact, $k$ is the Cartan matrix
of the Kac--Moody algebra extension of $G$.
\def\al#1{\alpha_{#1}}
Namely, $ 2k_{IJ}=(\al I\cdot\al J)/(\al I\cdot\al I)$,
where $\al0\equiv\sum_{i} \gamma^i\al i$.
Lax pair and B\"acklund
transformations are readily deduced from (4.6) and (4.24), (4.25).

 \bigskip\noindent
{\bf 5.  Super--Liouville Theory.}

Two--dimensional $N=1$ supergeometry is defined by
\footnote\dagger{
Einstein indices are denoted by $M, N, \cdots $, and run over
the coordinates $\xi, \bar \xi, \theta, \bar \theta$; frame
indices are denoted
by $ A = (a,\alpha)$, $a = z, \bar z$ and $\alpha = \pm$.  We have
$(\gamma^z)_{++} = (\gamma^{\bar z})_{--} = 1$, $\epsilon_z{}^z = -
\epsilon_{\bar z}{}^{\bar z} = i$ and $\epsilon_+{}^+ = -\epsilon_-{}^-
= i/2.$}
[18] a frame $E_M{}^A$ and a $U(1)$ connection $\Omega_M$.
The coordinates are denoted $(\xi,\overline\xi,\theta,\overline\theta)$
where $\xi,\overline\xi$ are commuting and $\theta,\overline\theta$
are anticommuting.
We will work with  functions
of these variables called ``superfields", which may be expanded in terms
of its anticommuting coordinates.
A superfield $F(\xi,\oxi,\theta,\oth)$, for instance, may
be expanded as
$$
F(\xi,\oxi,\theta,\oth)=f(\xi,\oxi)+\theta\psi(\xi,\oxi)+
\bar\theta \bar \psi(\xi,\oxi)+\theta\oth\hat f(\xi,\oxi) $$
Super--covariant derivatives on superfields of $U(1)$ weight $n$ are
defined by
$$
\d_A^{(n)} = E_A{}^M (\partial_M + i n \Omega_M)
\eqno(5.1)
$$
Torsion and curvature forms are defined by
$$
\eqalign{
T^A & = dE^A + E^B \wedge \Omega \epsilon_B{}^A = \half E^c \wedge E^B
T_{BC}{}^A\cr
R & = d\Omega = \half E^B \wedge E^A R_{AB} \cr}
\eqno(5.2)
$$
or in terms of structure relation:
$$
\left [\d_A, \d_B \right ]_\pm^{(n)} = T_{AB}{}^C \d_C + i n R_{AB}
\eqno(5.3)
$$
The standard $N=1$ supergravity torsion constraints are
$$
T_{ab}{}^c = T_{\alpha\beta}{}^\gamma = 0 \qquad\qquad T_{\alpha\beta}{}^c =
2(\gamma^c)_{\alpha\beta}
\eqno(5.4)
$$
These constraints are left invariant under super Weyl transformations
given by
$$
\eqalign{
E_M{}^a & = \exp (\Phi) \hat E_M{}^a\cr
E_M{}^\alpha & = \exp \left (\half \Phi\right )
\left ( \hat E_M{}^\alpha + \hat E_M{}^a (\gamma_a)^{\alpha\beta}
\d_\beta \Phi\right ) \cr
\Omega_M & = \hat \Omega_M + \hat E_M{}^a \epsilon_a{}^b \hat \d_b \Phi
+ \hat E_M{}^\alpha \epsilon_\alpha{}^\beta \hat\d_\beta \Phi\cr}
\eqno(5.5)
$$
In general torsion and curvature transform as
$$
T_a=\exp(\Phi)\hat T^a \qquad T^\alpha = \exp (\half \Phi)\hat T^\alpha \qquad
R_{+-} = \exp (-\Phi) (\hat R_{+-} - 2i \hat \d_+ \hat D_-\Phi)
\eqno(5.6)
$$
When $R_{+-}$ is set to a constant in (5.6), the second line becomes the
Liouville equation on the field $\Phi$, assuming that the torsion
constraints (5.4) are satisfied on the background geometry.

Topological supergravity is based on the gauge (super) group $OSp(1,1)$
[19].  It is convenient to rewrite its structure relations as
$$
[J_3, J_A] = -i \epsilon_A{}^B J_B;\quad [J_A, J_B]_\pm =
\Gamma_{AB}{}^C J_C + \Gamma_{AB}{}^3 J_3
\eqno(5.7)
$$
where the structure constants are given by
$$
\eqalign{
\Gamma_{\alpha\beta} {}^c
& = 2(\gamma^c)_{\alpha\beta};\quad \Gamma_{\alpha b}{}^\gamma = -
\Gamma_{b\alpha}{}^\gamma = \mu (\gamma_b)_\alpha{}^\gamma;\cr
\Gamma_{z\bar z}{}^3 & = - \Gamma_{\bar zz}{}^3 = - 2\mu^2 ;\quad
\Gamma_{+-}{}^3 = \Gamma_{-+}{}^3 = 2\mu\cr}
\eqno(5.8)
$$
all other components vanish.  We now introduce $OSp(1,1)$
valued gauge fields, decomposed onto frame and connection as follows:
$$
\c A = -i \Omega J_3 + E^A J_A
\eqno(5.9)
$$
The curvature form decomposes as follows:
$$
\c F =  \left (T^A + \half E^C \wedge E^B \Gamma_{BC}{}^A\right ) J_A
- i \left (R + {i\over 2} E^C\wedge E^B \Gamma_{BC}{}^3\right ) J_3
\eqno(5.10)
$$
 From (5.6), it is now clear that $\c F=0$ corresponds to the
super--Liouville equations with $R_{+-} = \mu$.

To construct the Lax pair, we let $OSp(1,1)$ act on a triplet $\psi =
(\psi_1,\psi_2,\psi_3)^T$ transforming under  the fundamental representation.
The equations
$$
\left ( \partial_M + \c A_M\right ) \Psi = 0
\eqno(5.11)
$$
are integrable precisely when $\c F=0$, which is just the super-Liouville
equation, plus the $N=1$ torsion constraints; thus (5.11) is the Lax
pair for this system.
To exhibit the super--Liouville field explicitly, we perform a super--Weyl
transformation (5.5), and we obtain the following projections onto frame
indices:
$$
\left (\hat \d_\alpha^{(0)} - i \hat\Omega_\alpha J_3 + J_\alpha \exp
\left ( \half\Phi \right ) - i(\gamma_5)_\alpha{}^\beta
\hat\d_\beta \Phi J_3\right ) \psi = 0\qquad\qquad\qquad
\eqno(5.12a)
$$
$$
\left (\hat \d_a^{(0)} - i \hat\Omega_a J_3 +\exp(\Phi) J_a +
(\gamma_a)^{\alpha\beta} \hat\d_\beta  \Phi \exp \left (
\half\Phi\right ) J_\alpha
-i\epsilon_a{}^b \hat\d_b \Phi J_3\right ) \psi = 0
\eqno(5.12b)
$$
By squaring the differential operator in (5.12a) and using the torsion
constraints, one precisely recovers (5.12b), so that we just retain
(5.12a) as the basic Lax pair.

To construct the B\"acklund transformation, we introduce the
inhomogeneous coordinates
$$
\exp \Sigma = {\psi_2\over \psi_1} \qquad\qquad \eta = {\psi_3 \over
\sqrt{\psi_1\psi_2}}
\eqno(5.13)
$$
assuming that $\psi_1 $ and $\psi_2$ are of even and $\psi_3$ of odd
grading; (5.12a) now decomposes as follows:
$$
\eqalign{
& \hat\d_+^{(0)} \Sigma + i \hat\Omega_+ - \hat\d_+\Phi =
\sqrt{\lambda_z} \eta \exp \left (\half \Phi + \half \Sigma \right
)\cr
& \hat\d_-^{(0)} \Sigma + i \hat\Omega_ - + \hat\d_-\Phi =
\sqrt{\lambda_{\bar z}} \eta \exp \left (\half \Phi - \half \Sigma \right
)\cr
&\hat\d_+^{(0)} \eta = - \sqrt{\lambda_z} \exp \left (\half \Phi +
\half \Sigma \right )\cr
&\hat\d_-^{(0)} \eta = - \sqrt{\lambda_{\bar z}} \exp \left (\half \Phi -
\half \Sigma \right )\cr}
\eqno(5.14)
$$
The system of equations (5.14) is integrable provided $\Phi$ satisfies
the super--Liouville equation, whereas $\Sigma$ satisfies a linear
equations, analogous to (2.13):
$$
2 i \hat\d_+^{(-1/2)} \hat\d_-^{(0)} \Sigma =
\hat\d_-^{(1/2)} \hat\Omega_+ - \hat \d_+^{(-1/2)} \hat\Omega_-
\eqno(5.15)
$$
Lax pair (5.12a) and B\"acklund transformation (5.14) generalize those
that were known on a background of two--dimensional flat space [20].  In
turn, we see that the $OSp(1,1)$ Toda system on the plane with global
$N=1$ supersymmetry may be consistently coupled to $N=1$ supergravity.

\bigskip

\noindent
{\bf 6. Toda Theory for Supergroups and $N=1$ Supergeometry.}

We shall now be interested in coupling Toda theory for supergroups $G$
[21] to Riemannian geometry and $N=1$ supergeometry. We begin by
considering a $G$--valued connection:
$$
\c A = \sum_i \Omega^i h_i + \sum_{\gamma\in\Delta} E^\gamma x_\gamma
\eqno(6.1)
$$
with curvature equations:
$$
\eqalign{
\c F & = \sum_i d\Omega^i h_i - \half \sum_{\gamma\in\Delta} E^{-
\gamma} \wedge E^\gamma [x_\gamma, x_{-\gamma}]\cr
\qquad& + \sum_{\gamma\in\Delta} \left [dE^\gamma +
\sum_{i,j} \Omega^i k_{ij} \gamma^j \wedge E^\gamma - \half
\sum_{\gamma=\gamma'+\gamma''}E^{\gamma'} \wedge E^{\gamma''}
N_{\gamma''\gamma'} \right ] x_\gamma\cr}
\eqno(6.2)
$$
We shall again postulate the equation $\c F=0$ for $G$--gravity.

Two classes of supergroups $G$ must be distinguished [22].  First, we
have supergroups for which {\bf all} simple positive roots may be chosen
to have odd grading.  In the Kac classification, the finite dimensional
group with this property are $A(n,n-1), B(n-1,n), B(n,n), D(n+1,n),
D(n,n)$ and $D(2,1;\alpha)$.  For these groups, Toda field theory in
flat space is $N=1$ supersymmetric and we shall see that in these cases,
Toda field theory may be coupled to $N=1$ supergravity in an invariant
way.

The second class of supergroups are those for which at least one simple
root must have even grading, \ie\ all the others.  Corresponding Toda
field theories in flat space are not $N=1$ supersymmetric, and they
cannot be coupled to $N=1$ supergravity in an invariant way.  We shall
couple them here only to ordinary gravity.

For supergroups in the first group, we use the convention that all
simple roots are of odd grading and
$$
N_{{\alpha}_{i},\alpha_{j}} = N_{{-\alpha}_{i},{-\alpha_{j}}} = 2
$$
We recover $N=1$ super Toda theory by setting

$$
\eqalign{
E^{\alpha_i} & = \exp \half \Phi_i \left [E^+ + (\d_+\Phi_i)E^z\right ]\cr
E^{-\alpha_i} & = \exp \half \Phi_i \left [E^- + (\d_-\Phi_i)E^{\bar
z}\right ]\cr
E^{\alpha_i + \alpha_j}&  = \exp\half (\Phi_i+\Phi_j)E^z
\times\cases{2 & $i\not= j$\cr
1 & $i=j$\cr} \cr
E^{- \alpha_i - \alpha_j}&  = \exp\half (\Phi_i+\Phi_j)E^{\bar z}
\times\cases{2 & $i\not= j$\cr 1& $i=j$\cr} \cr
E^\gamma & = 0 {\rm\ if\ } |\eta(\gamma)| \geq 3.\cr}
\eqno(6.3)
$$
and where $\left (E^A, \Omega\right )$ is an arbitrary two dimensional
supergeometry defined in (5.1).  The torsion type equations are solved by
$$
i \sum_j \Omega^j k_{ji} = \half \Omega \rho_i + E^A J_A{}^B \d_B\Phi_i
\eqno(6.4)
$$
with the supercomplex structure $J_A{}^B$ defined by
$$
J_A{}^B = \delta_A{}^B \times \cases{+i & $A = z,+$\cr
        -i & $A = \bar z,-$\cr}
\eqno(6.5)
$$
The flatness condition $\f_{\pm\pm} = 0$ evaluated on (6.3) and (6.4)
reduces to the torsion constraints of ordinary supergravity and $\f_{+
-} = 0$ yields the Toda equation, coupled to a general $N=1$ supergravity
background:
$$
\d_-\d_+ \Phi_i + \sum_j \exp (\Phi_j) k_{ji} - {i\over 2} {\cal R_{+-}}
\rho_i = 0
\eqno(6.6)
$$
The other components of $\c F_{AB}=0$ are automatically
satisfied using this Toda equation, the Jacobi identity and
the torsion constraints.
The Lax pair is easily written down in terms of $\pm$ components:
$$
\eqalign{
& \left ( \d_+^{(0)} + \sum_i \Omega_+^i h_i +\sum_i \exp \left (\half
\Phi_i\right ) x_{\alpha_{i}} \right ) \Psi = 0\cr
& \left ( \d_-^{(0)} + \sum_i \Omega_-^i h_i + \sum_i\exp \left (\half
\Phi_i\right ) x_{-\alpha_{i}} \right ) \Psi = 0\cr}
 \eqno(6.7)
$$
Notice that these supergroups always contain the basic supergroup
$ OSp(1,1)$.

In the case of supergroups for which all simple roots cannot be chosen
odd grading, we may couple Toda field theory only to ordinary gravity.
In this case, it is still convenient to use the notation of supergravity
and we embed gravity into supergravity by setting the gravitino and
auxiliary fields to zero in Wess--Zumino gauge:
$$
 \d_+^{(n)} = \partial_\theta + \theta D_z^{(n)} - {i\over 2}
\theta\bar\theta \omega_z \partial_{\bar\theta};\qquad
 \d_-^{(n)} = \partial_{\bar\theta} + \bar\theta D_{\bar z}^{(n)} -
{i\over 2} \theta\bar\theta \omega_{\bar z} \partial_\theta
\eqno(6.8)
$$
The gauge field is of the following form
$$
\eqalign{
\c A_+ & = \sum_i \Omega_+^i h_i + \sum_{\alpha_i\in\Delta_s^{\rm
odd} } \exp \left( \half \Phi_i\right) x_{\alpha_{i}}
+ \sum_{\alpha_{i}\in\Delta_s^{\rm even}}
\theta \exp \left(\half \Phi_i\right) x_{\alpha_{i}}\cr
\c A_- & = \sum_i \Omega_-^i h_i + \sum_{\alpha_i\in \Delta_s^{\rm
odd} } \exp \left( \half \Phi_i\right) x_{-\alpha_{i}}
+ \sum_{\alpha_{i}\in \Delta_s^{\rm even}} \bar\theta \exp \left(
\half \Phi_i\right) x_{-\alpha_{i}} \cr}
\eqno(6.9)
$$
Where $\Delta_s^{\rm odd}$, $\Delta_s^{\rm even}$ denote the odd and
even simple roots respectively.
Analogous to  the previous case,
the zero curvature condition reduces to the Toda equation coupled
to gravity
$$\c D_-\c D _+\Phi_i
+\sum_{\al j\in\oddroots} \exp \{\Phi_j\} k_{ji}
+\sum_{\al j\in\evenroots} \theta\bar \theta\exp \{ \Phi_j\}k_{ji}
-{i\over2}R_{+-} \rho _i =0\eqno(6.10)$$
\endpage
\bigskip \noindent{\bf References}
\item{[1]} C. Gardner, J. Greene, M. Kruskal and R. Miura, \prl
{\bf 19} (1967) 1095\nextline
         P. D. Lax, {\sl Comm. Pure. Appl. Math.} {\bf 21} (1968)
        647.\nextline
        V. E. Zakharov and L. Faddeev, {\sl Funct. Anal. Appl.} {\bf 5}
        (1971) 18\nextline
        V. E. Zakharov and A. B. Shabat, {\sl Funct. Anal. Appl. }{\bf 8}
        (1974) 43\nextline
        L. D. Faddeev and L. A. Takhtajan, {\sl``Hamiltonian Methods in the
Theory of Solitons''}, Springer, Berlin 1987.\nextline
        A. Dold and B. Eckmann, ed. Lecture Notes in Mathematics
        {\sl ``B\"acklund Transformations''}, No. 515, Springer, 1974.\nl
M.J. Ablowitz, H. Segur, {\sl``Solitons and the Inverse Scattering Transform"},
  Siam, Philadelphia 1981.

\item{[2]} E. Witten, \prl{\bf 38} (1977) 121\nl
  M.F.~Atiyah, R.S.~Ward, \cmp{\bf55} (1977) 124

\item{[3]}J. Liouville, {\sl J. Math Pure Appl.} {\bf 18} (1853) 71\nextline
        E. D'Hoker and R. Jackiw, \prd{\bf  D26} (1982) 3517\nextline
        E. Braaten, T. Cutright, G. Ghandour and C. Thorn, \anp
        (N.Y.) {\sl147} (1983) 365.

\item{[4]} E. D'Hoker, \plb {\bf 264B} (1991) 101.

\item{[5]} L. Castellani, R. D'Auria and P. Fr\'e, in ``Supersymmetry
and Supergravity'', Proc. XIX--th Winter School and Workshop on
Theoretical Physics (Karpacz, 1983) ed. B. Milewski, World Scientific
Publ.

\item{[6]}R. Jackiw in ``Quantum Theory of Gravity'', ed. S.
Christensen, Adam Hilger, Bristol, 1984. \nextline
        C. Teitelboim, {\it ibid.} \nextline
        K. Isler and C. Trugenberger, \prl {\bf 63} (1989)
        834.\nextline
        A. H. Chamseddine and D. Wyler, \plb{\bf 228B} (1989) 75.

\item{[7]}J. M. F. Labastida, M. Pernici and E. Witten, \npb {\bf
        B310} (1988) 611.\nextline
        D. Montano and J. Sonnenschein, \npb {\bf B313} (1990) 258;
{\bf B324} (1989) 348.\nextline
        R. Myers and V. Periwal, \npb{\bf B333} (1990) 536.\nextline
        E. Witten, \npb{\bf B340} (1990) 281\nextline
        R. Dijkgraaf and E. Witten, \npb{\bf B342} (1990) 486.

\item{[8]}C. M. Hull, \npb{\bf B364} (1991) 621; \plb{\bf B269}
(1991) 257.\nextline
        C. N. Pope, {\sl``Lectures on W--Algebras and W--gravity''},
        CTP--TAMU--103--91 (1991)\nextline
        A. Gerasimov, A. Levin and A. Marshakov, \npb{\bf B360}(1991) 537.

\item{[9]}K. Aoki and E. D'Hoker, \npb{\bf B387} (1992) 576

\item{[10]}S. Kobayashi and K. Nomizu, {\sl``Foundations of Differential
Geometry''}, Vol. I (Wiley, New York, 1963).\nextline
        S. Helgason, {\sl``Differential Geometry, Lie Group and Symmetric
        Spaces''}, Academic Press (1978).

\item{[11]} D. Cangemi, R. Jackiw, \prl{\bf 69} (1992) 233

\item{[12]} C.G. Callan, S.B. Giddings,
J.A. Harvey, A. Strominger, \prd{\bf D45} (1992) 1005

\item{[13]}K. Li, Nucl. Phys. {\bf B346} (1990) 329.\nextline
        K. Schoutens, A. Sevrin and P. van Nieuwenhuizen, \ijp
        {\bf A6} (1991) 2891.

\item{[14]}     J.-L Gervais and Y. Matsuo,
  LPTENS--91/35 (1991), \plb{\bf 274B} (1992) 309

\item{[15]}A. N. Leznov and M. V. Saveliev, \lmp {\bf 3}
        (1979) 207.\nextline
        P. Mansfield, \npb{\bf B208} (1982) 277.

\item{[16]}M. V. Saveliev, \plb {\bf A122} (1987) 312.

\item{[17]} G.V. Dunne, R. Jackiw, S-Y. Pi, C.A.~Trugenberger,
\prd{\bf D43} (1991) 1332

\item{[18]}P. Howe, {\sl J. Phys. }{\bf A12} (1979) 393.\nextline
        J. Wess and J. Bagger, {\sl``Supersymmetry and Supergravity''},
        Princeton (1983)\nextline
        E. D'Hoker and D. H. Phong, \rmp {\bf 60} (1988) 17.

\item{[19]}D. Montano, K. Aoki and J. Sonnenschein, \plb {\bf
B247} (1990) 64.

\item{[20]}E. D'Hoker, \prd {\bf D28} (1983) 1346.

\item{[21]}M. A. Olshanetsky, \cmp {\bf 88} (1983)
        1205.\nextline
        J. Evans and T. Hollowood, \npb{\bf B352} (1991) 723.\nextline
        T. Inami and H. Kanno, \cmp {\bf 136} (1991) 543.

\item{[22]}J. E. Humphreys, {\sl``Introduction to Lie Algebras and
Representation Theory''}, Springer--Verlag (1972)\nextline
        V. G. Kac, {\sl``Infinite Dimensional Lie Algebras''}, Cambridge
        University Press (1985)\nextline
        V. G. Kac, {\sl Adv. Math} {\bf 26} (1977) 8\nextline
        D. A. Leites, M. V. Saveliev, V. V. Serganova, in {\sl ``Group
        Theoretical Methods in Physics''}, VNU Science Press (1986).
\endpage
\end